\documentclass[5p,twocolumn]{elsarticle} 
\usepackage{graphicx}
\usepackage{epstopdf}

\begin{document}

\oddsidemargin = -12pt

\title{Quartet excitations and cluster spectra in light nuclei}

\author{
J. Cseh  and G. Riczu\\
Institute for Nuclear Research, Hungarian Academy of Sciences,
                Debrecen, Pf. 51, Hungary-4001} 
 
\date{\today}

%\maketitle
\begin{abstract}
The relation of quarteting and clustering in atomic nuclei is discussed
based on symmetry-considerations. This connection enables us to predict a complete
high-energy cluster spectrum from the description of the low-energy
quartet part. As an example the $^{28}$Si nucleus is considered,
including its well-established ground-state region, the recently proposed
superdeformed band, and the high-lying molecular resonances.
\end{abstract}

\begin{keyword}
quarteting and clustering, excitation spectrum, multichannel dynamical symmetry
\PACS 21.60.Fw, 21.60.Cs, 27.30.+t
\end{keyword}

\maketitle

\oddsidemargin = -12pt

Most of the atomic nuclei are typical mesoscopic systems, which allow neither
ab initio, nor statistical description. Therefore, models play the crucial role
in the understanding the nuclear structure. The fundamental structure models are based 
on different physical pictures, e.g. shell, cluster or liquid drop, therefore, their
interrelation is not trivial. Symmetry-considerations are very helpful in finding
their connection, as well as in describing complex spectra. In this letter we show
how the nucleon-quarteting, which is a shell model phenomenon, is related to
the clusterization, i.e. to the appearance of a molecule-like configuration.
We do so by applying a semimicroscopic algebraic description for both phenomena,
which reveals a special symmetry, called multichannel dynamical symmetry.
This symmetry allows us to obtain a high-lying cluster spectrum from the quartet model
fitted to the low-energy part. We do not know any other method of this
ability. 

The investigation of quarteting and clustering has a long history, and a large variety 
of models have been invented for their description. 
When the cluster is an alpha-particle, which is the most typical and best
studied case, the two structures are obviously related to each other:
in both cases the basic building block is composed of two protons and
two neutrons. 
In the phenomenological approaches, which do not respect the Pauli-exclusion
principle, the wavefunction of the shell-like and molecule-like configurations
(or those of two different cluster configurations) are orthogonal to each other.
In fact, however, the antisymmetrization modifies the simple geometric
picture, and as a result, the overlap can be finite, up to 100 percent.
One needs microscopically constructed model spaces for the study of
this connection. (Whether the interactions are also microscopic
or not, i.e. if the description is fully microscopic, or
semimicroscopic is less relevant in this respect.)

In what follows we apply semimicroscopic algebraic models for the
description of both quarteting and clustering. This approach takes
into account the exclusion principle, furthermore, due to its fully
algebraic nature it has rather
transparent symmetry properties.
(We call a model fully algebraic when 
not only the basis states, but the physical operators
as well are characterized by group representations.)
\\

\noindent
{\it 
The  semimicroscopic algebraic quartet model} 
(SAQM)
\cite{quart}
is a symmetry-governed truncation of the no-core shell model
\cite{nocore},
that describes the quartet excitations in a nucleus.
A quartet is formed by two protons and two neutrons,
which interact with each other very strongly,
as a consequence of the short-range attractive forces 
between the nucleons inside a nucleus
\cite{arima}.
The interaction between the different quartets is weaker.
In this approach 
the L-S coupling is applied, the model space has a spin-isospin sector,
characterized by Wigner's U$^{ST}$(4) group
\cite{wigner},
and a space part described by Elliott's U(3)
\cite{elliott}.
Four nucleons form a quartet  
\cite{harvey}
when their
spin-isospin symmetry is \{1,1,1,1\}, and their permutational symmetry is \{4\}. 
This definition allows two protons and two neutrons to form a
quartet even if they sit in different shells. As a consequence
the quartet model space incorporates 0, 1, 2, 3, 4, ... major shell
excitations (in the language of the shell model),
contrary to the original interpretation of
\cite{arima},
when the four nucleons had to occupy the same single-particle
orbital, therefore, only 0, 4, 8, ... major shell excitations could be
described. 

The model is fully algebraic,
% in the sense that not only its basis states
%are characterized by group symmetries, but also 
%the physical operators are written in terms of group generators
%\cite{quart}.
therefore, group theoretical methods can be applied in calculating the matrix elements.
The operators contain parameters to fit to the experimental data,
that is why the model is called semimicroscopic: phenomenologic
operators are combined with microscopic model space.
Due to the quartet symmetry only a single \{1,1,1,1\} U$^{ST}$(4) sector plays a 
role in the calculation of the physical quantities,
thus the U(3) space-group and its subgroups are sufficient for characterizing the situation:
\begin{eqnarray}
 U(3) \supset SU(3) \supset SO(3) \supset SO(2)
 \nonumber \\
\vert [n_1,n_2,n_3]  ,  (\lambda , \mu) ,\  K\ ,  \ \ L \ \ \ \ ,\ \ \  M  \ \rangle .
 \label{eq:ellgrch}
\end{eqnarray}
In Eq. (1) we have indicated also the representation labels of the groups
which serve as quantum numbers of the basis states.  
Here $ n= n_1 + n_2 +n_3 $ is the number of the oscillator quanta, and
 $\lambda = n_1 - n_2, $
 $\mu = n_2 -n_3 $.
The angular momentum content of a $(\lambda ,\mu )$ representation is as
follows 
\cite {elliott}: 
$L= K, K + 1,...,K + max {(\lambda , \mu)}$, 
$ K = min {(\lambda , \mu )},
 min {( \lambda , \mu )}  - 2,..., 1 \ or \ 0,$
%\begin{equation}
 %\begin{array}{l}
 %\displaystyle{
 %L = K_L, K_L + 1,...,K_L + max {(\lambda , \mu)},
 %\label{mlett:1}
 %} \\  \ \\
 %\displaystyle{
 %K_L = min {(\lambda , \mu )},
 %min {( \lambda , \mu )}  - 2,..., 1 \ or \ 0,
 %} \end{array}
%\eqnum{4.2} 
%\end{equation}
with the exception of $K_L = 0$, for which
%\begin{equation}
$ L = max {(\lambda , \mu)}, max {(\lambda , \mu)} - 2,..., 1 \ or \ 0 $.
%\eqnum{4.3}
In the limiting case of the dynamical symmetry, when the Hamiltonian
is expressed in terms of the invariant operators of this group-chain,
an analytical solution is available for the energy-eigenvalue problem
(an example is shown below).

The SAQM can be considered as an effective model in the sense of
\cite{effective}:
the bands of different quadrupole shapes are described by their lowest-grade
U(3) irreducible reperesentations
(irreps) without taking into account the giant-resonance excitations, built
upon them, and the model parameters are renormalised for the subspace of the
lowest U(3) irreps. 
\\

\noindent
{\it 
The semimicroscopic algebraic cluster model}
 (SACM) 
\cite{sacm},
just like  the other cluster models,
classifies the relevant degrees of freedom of the nucleus  into  two categories:
they belong either to the internal structure of the clusters, or to their relative motion.
In other words:  the description is based on a molecule-like picture.
The internal structure of the clusters is handled in terms of
Elliott's shell model
\cite{elliott} with
U$^{ST}$(4)$\otimes$U(3) group structure
(as discussed beforehand).
The relative motion is taken care of by the vibron model
\cite{vibron},
which is an algebraic model of the dipole motion, and
it has a U(3) basis, too.
For a two-cluster-configuration this model has a group-structure of
U$^{ST}_{C_1}$(4)$\otimes$U$_{C_1}$(3) $\otimes$
U$^{ST}_{C_2}$(4)$\otimes$U$_{C_2}$(3) $\otimes$
U$_R$(4).

The model space is constructed also in this case in a microscopic way,
i.e. the Pauli-forbidden states are excluded. It requires the truncation
of the basis of the vibron model, as given by the Wildermuth-condition
(see below for some  specific examples). This condition determines the
lowest-allowed quantum number of the relative motion, i.e.
the allowed major shells of the (united) nucleus. Furthermore,
one needs to distinguish between the Pauli-allowed and forbidden
states within a major shell, too. Different methods can be applied
to this purpose; e.g. by making an intersection with the U(3) shell
model basis of the nucleus, which is constructed to be free from
the forbidden states.
The SACM is fully algebraic, and semimicroscopic in the sense
discussed above.

When we are interested only in spin-isospin zero states of the nucleus
(a typical problem in cluster studies, and being our case here, too),
then only the space symmetries are relevant (apart from the construction
of the model space).
Considering, for the sake of simplicity, a binary cluster configuration
the corresponding group-chain is:    
\begin{eqnarray}
 U_{C_1}(3)  \otimes  U_{C_2}(3)  \otimes \  U_R(4)  \supset 
 U_C(3)  \otimes U_R(3)  \supset 
 \nonumber
 \\
 %\supset   
 U(3)  \supset SU(3) \supset SO(3)  \supset SO(2) .
\end{eqnarray}
The basis defined by this chain is especially useful
for treating the exclusion principle, since the U(3) generators commute
with those of the permutation group, therefore, all the basis states of an 
irrep are either Pauli-allowed, or forbidden
\cite{horisup}.           
In particular, this U(3) basis allows us to pick up the allowed cluster
states from the U(3) shell model basis (1).         

A Hamiltonian corresponding to the dynamical symmetry of group-chain (2) reads as:      
\begin{eqnarray}
{\hat H} =
{\hat H_{C_1}} + {\hat H_{C_2}} + {\hat H_{U_R(4)}} +
{\hat H_{U_C(3)}} + {\hat H_{U_R(3)}} +
 \nonumber
\\
{\hat H_{U(3)}} + {\hat H_{SU(3)}} + {\hat H_{SO(3)}} .
\end{eqnarray}
We note here, that the first part 
\begin{equation}
{\hat H_{CM}} =
{\hat H_{C_1}} + {\hat H_{C_2}} + {\hat H_{U_R(4)}} +
{\hat H_{U_C(3)}} + {\hat H_{U_R(3)}} 
\end{equation}
is an operator that corresponds to  the pure cluster picture, while the second part
\begin{equation}
{\hat H_{SM}} =
{\hat H_{U(3)}} + {\hat H_{SU(3)}} + {\hat H_{SO(3)}} 
\end{equation}
is a shell model Hamiltonian (of the united nucleus). 
\\

\noindent
{\it 
The multichannel dynamical symmetry}
(MUSY)
\cite{musy1,musy2} 
connects different cluster configurations (including the shell model limit)
in a nucleus.
Here the word channel refers to the reaction channel,
that defines the cluster configuration.

The simplest case is a two-channel symmetry connecting two different
clusterizations. It holds, when both cluster configurations can be described
by an U(3) dynamical symmetry and in addition a further symmetry connects 
them to each other. This latter symmetry is that of the Talmi-Moshinsky
transformation.
It acts in the
pseudo space of the particle indices, or geometrically it corresponds to the
transformations between the different sets of Jacobi-coordinates associated
to the cluster configurations
\cite{hori,musy2}. 
The   ${\hat H_{SM}}$ Hamiltonian of Eq. (5)
is symmetric with respect to these transformations, therefore, it is invariant
under the changes from one clusterization to the other.
The cluster part of the Hamiltonian 
${\hat H_{CM}}$
is affected by the transformation from one configuration to the other,
of course. Nevertheless, it may remain invariant, which is the case for
simple operators, like the harmonic oscillator Hamiltonian, or
the quadrupole operator
\cite{musy2}.
Due to this symmetry of the quadrupole operator, the $E2$ transitions
of different clusterizations also coincide, when the MUSY holds,
just like the energy eigenvalues of the symmetric Hamiltonians
\cite{musy2}.

The MUSY is a composite symmetry of a composite system.
Its logical structure is somewhat similar to that of the dynamical supersymmetry
(SUSY) of nuclear spectroscopy. In the SUSY case the system has two
components, a bosonic and a fermionic one, each of them showing a
dynamical symmetry, and a further symmetry connects them to each other.
The connecting symmetry is that of   the supertransformations 
which change  bosons into  fermions or vice versa. 
In the MUSY case the system has two (or more) different clusterizations,
each of them having  dynamical symmetries which are connected to each other
by the symmetry of the (Talmi-Moshinsky) transformations that change
from one configuration to the other.     
               
When the multichannel dynamical symmetry  holds then the spectra
of different clusterizations are related to each other by very strong constraints.
The MUSY provides us with a unified multiplet structure of different cluster configurations,
furthermore the corresponding energies and E2
transitions coincide exactly. 
Of course, it can not be decided a priori whether the MUSY holds or not,
rather one can suppose the symmetry and compare its consequences
with the experimental data.
In what follows we
derive the spectra of two clusterizations from the quartet spectrum of the 
$^{28}$Si nucleus.
\\

\noindent
{\it 
The $^{28}$Si nucleus}
provides us with many reasons to be chosen  as
an illustrative example.
i) It has a well-established band-structure in the low-energy region, and to several
bands SU(3) quantum numbers could be associated as a joint conclusion of
experimental and theoretical investigations
\cite{sisu3}.
ii) More recently a new candidate was proposed for the superdeformed (SD)
band 
%from the combination of the experimental arguments from previous and recent measurements
\cite{jenkins}. 
Theoretical studies predicted the SD band 
\cite{amd,simi}
in line with the experimental observation.
iii) There are two cluster configurations:
$^{24}$Mg+$^{4}$He, and
$^{16}$O+$^{12}$C,
belonging to reaction channels in which fine-resolution measurements
revealed a rich spectrum of resonances.

In  
\cite{musy1}
the connection of these two cluster configurations has been discussed in terms
of the multichannel dynamical symmetry. In the present work we go beyond the former description in several aspects.
We calculate the quartet spectrum of the $^{28}$Si nucleus, and obtain the spectra of both clusterizations from the quartet excitations by projection,  without fitting anything to the cluster states,
i.e. the cluster spectra appear as  pure predictions.
In doing so we apply a simple
Hamiltonian with less number of parameters than in
\cite{musy1}.
In addition to the energy spectra we give the E2 transition ratios as well.
The new  superdeformed candidate band is also taken into account.
\\

\noindent
{\it Quartet excitations.}
%In the semimicroscopic algebraic quartet model the description of the quartets is
%based on the nucleonic degrees of freedom
%\cite{quart}.
%i.e. on the microscopic level. 
%A quartet is composed of two protons and two neutrons with a spin-isospin
%symmetry of \{1,1,1,1\}, thus only this section of the shell model space 
%needs to be taken into account. Also the physical operators have spin-isospin zero
%nature. Therefore, the basis states of the model are characterised by
%the representation labels of Elliott's U(3) group and its subgroups.
%The Hamiltonian and the transition operators are  expressed (in the
%simplest approximation) in terms of U(3) generators.

The lower most part of Figure 1 shows the experimental bands of the $^{28}$Si
nucleus, as established in 
\cite{sisu3}
together with the recently found superdeformed (SD) band 
\cite{jenkins}.
An especially favourable circumstance is that SU(3) quantum numbers are
associated to several experimental bands, 
%as a joint conclusion of the previous experimental and theoretical investigations, 
without any reference
to the quartet or cluster studies.
(In the experimental spectrum $\beta$ means $\beta$-instabil, while O and P
stand for oblate and prolate, respectively.)

\begin{figure}[placement !]
\includegraphics[height=18.1cm,angle=0.]{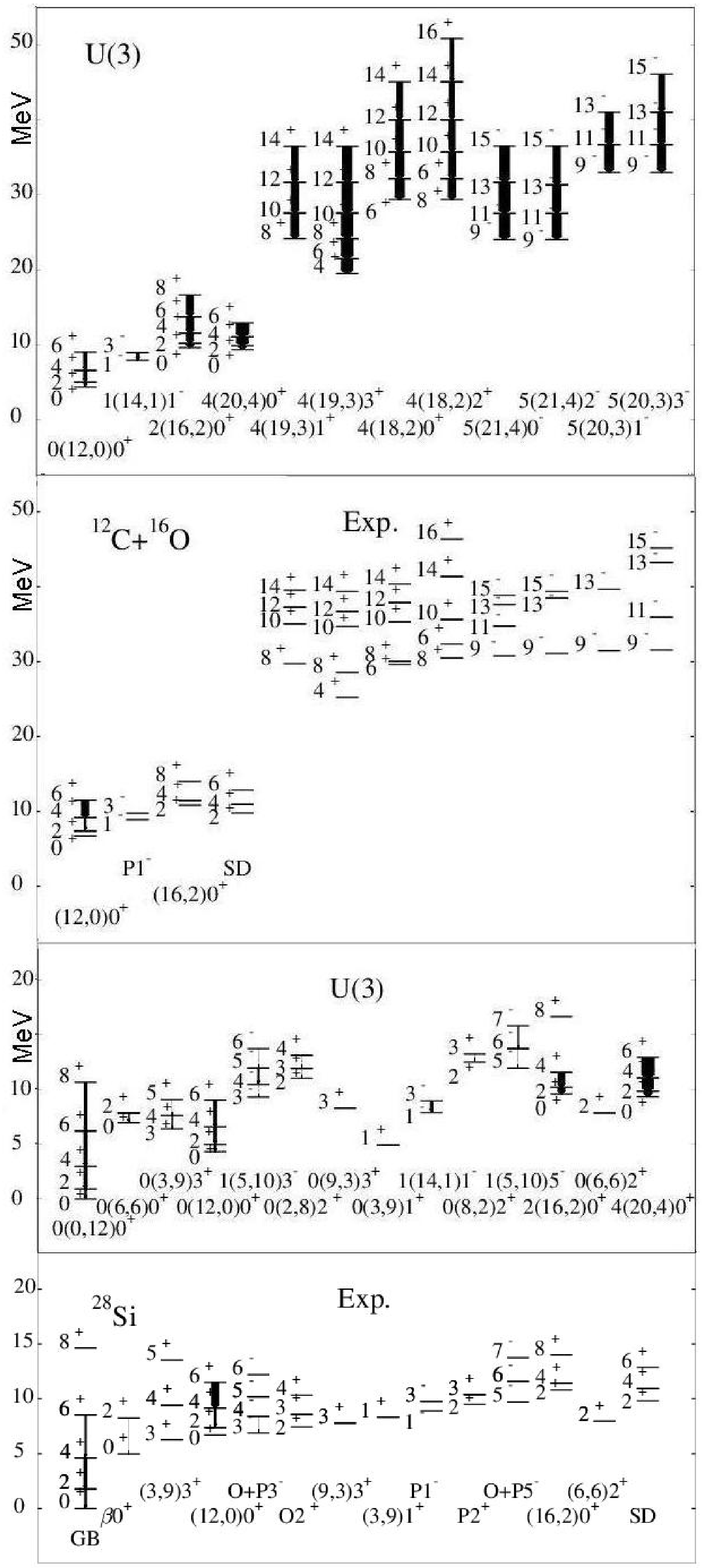}
\caption{ 
The spectrum of the semimicroscopic algebraic quartet model in
comparison with the experimental data of the  $^{28}$Si nucleus (lower part).
The experimental bands are labeled by the available quantum numbers, and the 
model states by the $n (\lambda, \mu)K^{\pi}$ labels. 
%The spin-parity in parenthesis indicates uncertain band-assignment.
The width of the arrow between the states is proportional to the
strength of the $E2$ transition. 
The upper part shows the $^{12}$C+$^{16}$O cluster spectrum,
which is obtained as a projection from the quartet spectrum,
without any further fitting.
\label{fig:spectrum}}
\end{figure}

The U(3) spectrum (the second one from below in Figure 1)  
is calculated within the SAQM approach
\cite{quart}. 
The experimental states are described by the lowest-lying model bands
with the appropriate spin-parity content.
We have applied a U(3) dynamically symmetric Hamiltonian, i.e.
an operator expressed in terms of the invariant operators of
the group-chain: 
U(3) $\supset$ SU(3)  $\supset$ SO(3):
\begin{equation}
{\hat H} =
(\hbar \omega) {\hat n}  +
a{\hat C}^{(2)}_{SU3} +
b{\hat C}^{(3)}_{SU3} +
d {1 \over {2\theta}}{\hat L}^2.
\end{equation}
The first term is the harmonic oscillator Hamiltonian
(linear invariant of the U(3)), with a strength obtained from
the systematics
\cite{moli}
$\hbar \omega$ = $45 A^{-{1 \over 3}} -25  A^{-{2 \over 3}}$ MeV = 12.11 MeV.
%($n$  is the number of excitation quanta compared to the ground state.)
The second order  invariant of the SU(3) (${\hat C}^{(2)}_{SU3}$)
represents the  
quadrupole-quadrupole interaction,
while the  third order Casimir-operator 
(${\hat C}^{(3)}_{SU3}$)
distinguishes  between the prolate and oblate shapes. 
$\theta$ is the moment of inertia calculated classically for the rigid shape
determined by the U(3) quantum numbers (for a rotor with axial symmetry)
\cite{arxiv},
and the $a$, $b$ and  $d$ parameters were fitted to the experimental data:
$ a = -0.133$ MeV,
$b =0.000444$ MeV
$d = 1.003$ MeV.
The $B(E2)$ value is given as
\cite{quart}:
%\begin{widetext}
\begin{eqnarray}
%\begin{equation}
B(E2, I_i \rightarrow I_f) =
\ \ \ \ \ \ \ \ \ \ \ \ \ \ \ \ \ \ \ \ \ \ \ \ \ \ \ \ \ \ \ \ \ \ \ \ \ \ \ \ \ 
%\ \ \
% \ \ \ \ 
\nonumber \\
{{2I_f +1} \over {2I_i +1}} {\alpha}^2
\vert \langle  (\lambda , \mu) K I_i , (11)2 \vert \vert (\lambda , \mu) K I_f \rangle  \vert ^2
 C(\lambda , \mu) ,
\label{e2matrix}
\end{eqnarray}
%\end{equation}
%\end{widetext}
where
$
\langle  (\lambda , \mu) K I_i , (11)2 \vert \vert (\lambda , \mu) K I_f \rangle
$
is the SU(3) $\supset $ SO(3) Wigner coefficient,
%\cite{prog}, 
and $\alpha^2$ (= 0.366 W.u.) is a parameter
fitted to the   experimental value of the
$2^+_1 \rightarrow 0^+_1$ transition of 13.2 W.u. 
%(Further formulae for the transition rates can be found in e.g.
%\cite{bijkot,levcse}.)
%The interband transition rate is zero.
\\

\noindent
{\it Cluster spectra.}

The MUSY can connect the quartet (shell) model state
%(as a special cluster configuration)
 to other clusterizations, too.
Here we show, how the 
$^{24}$Mg+$^{4}$He, and
$^{16}$O+$^{12}$C
cluster spectra can be obtained from the quartet spectrum by simple projections.

In this description the clusters are considered to be in their intrinsic ground states.
Each of the four clusters of the present study have spin-isospin zero quantum 
numbers, i.e. they belong to 
scalar representations of Wigner's U$^{ST}$(4) group. Their space symmetry is
given by Elliott's U(3) group, which is known to be approximately valid for these
light nuclei, therefore,  simple leading representation characterize their
ground states as follows:
$^{4}$He: \{0,0,0\},
$^{12}$C: \{4,4,0\},
$^{16}$O: \{4,4,4\},
$^{24}$Mg:\{16,8,4\}. 

A state of an $\{n_1,n_2,n_3\}$ symmetry is present in a binary
cluster configuration $C_1 + C_2$, if the triple product matches with it:
\begin{eqnarray}
\{n^{c_1}_1,n^{c_1}_2,n^{c_1}_3\} \otimes
\{n^{c_2}_1,n^{c_2}_2,n^{c_2}_3\} \otimes
\{n_R,0,0\}  =
\nonumber\\
\{n_1,n_2,n_3\}  \oplus ...
\end{eqnarray}
where $\{n_R,0,0\}$ stands for the relative motion, and $n_R$ is limited from below,
due to the Pauli-principle (known as the Wildermuth-condition
\cite{wildkan}). 

In case of the 
$^{24}$Mg + $^{4}$He  clusterization the lowest allowed value of
$n_R$ is 8, showing that in the unification of the two nuclei the 
4 nucleons of the $^{4}$He has to be lifted to the 
2  $\hbar \omega $
major shell in order not to violate the exclusion principle.
For the 
$^{16}$O + $^{12}$C
clusterization the values below 16 are excluded.

We note here that in our case the results of the triple product
have always single multiplicity. This is because one of the clusters
has a closed-shell structure, i.e. it is an U(3) scalar.
As a consequence a single U(3) irrep is multiplied by the single-row
irrep of the relative motion $\{n_R,0,0\}$.

For illustration we show in Table I. the SU(3) quantum numbers
of the quartet model as well as the two cluster model spaces 
for 0  $\hbar \omega $.

\begin{table}
\caption{
SU(3)  quantum numbers  of the 0  $\hbar \omega $ states in the $^{28}$Si nucleus.
The superscripts indicate multiplicity.
}
\begin{tabular}{|c|c|c|c|}
\hline
\hline
%\vskip 4pt
$\hbar \omega$&Quartet&$^{24}Mg$+$\alpha$&$^{12}C$+$^{16}O$\\ \hline
     &$(12,0)^{1}$,$(0,12)^{1}$,      &$(12,0)^{1}$,$(0,12)^{1}$,    &$(12,0)^{1} $\\
     &$(3,9)^{1}$,$(9,3)^{1}$,          &$(3,9)^{1}$,$(9,3)^{1}$,        &\\
     &$(6,6)^{1}$,  $(2,8)^{2}$,        &$(6,6)^{1}$,$(2,8)^{1}$,        &\\
    &$(8,2)^{2}$,  $(5,5)^{2}$,        &$(8,2)^{1}$,$(5,5)^{1}$,        &\\
     &$(3,6)^{2}$,$(6,3)^{2}$,          &$(3,6)^{1}$,$(6,3)^{1}$,        &\\
0   &$(1,7)^{1}$,  $(7,1)^{1}$,        &$(4,4)^{1}$                               &\\
     &$(4,4)^{4}$,  $(2,5)^{1}$,        &                                                   &\\
     &$(5,2)^{1}$,$(0,6)^{3}$,          &                                                   &\\
     &$(6,0)^{3}$,  $(3,3)^{3}$,        &                                                   &\\
     &$(1,4)^{1}$,  $(4,1)^{1}$,        &                                                   &\\
     &$(2,2)^{3}$,$(0,0)^{2}$           &                                                   &\\   \hline
\hline
\end{tabular}
\end{table}

Until the basis states are determined by the U(3) (and its subgroups) symmetry,
and the interactions are dynamically symmetric, i.e. the MUSY holds, the corresponding
energies and E2 transition rates in the quartet  and cluster descriptions coincide. 
%(This is a 
%consequence of the fact that the U(3) generators are scalars with respect to the
%transformations in the particle index pseudo space.)  
Therefore, by applying the 
selection rule (8), not only the  cluster model basis states, but also their energy
eigenvalues, as well as the E2 transition probabilities between them 
can be selected. In other words the cluster spectrum is obtained from
the quartet one by a simple projection.

The $^{12}$C+$^{16}$O spectrum of Figure 1 shows those bands of
the low-energy part, which are present in this cluster configuration,
as well as the resonance spectrum from the heavy ion experiments,
according to the compilation of
\cite{abon}.
The latter one is  organised into bands according to their energy-differences.
The corresponding U(3) spectrum is calculated with Eq. (6), without
fitting anything to the high-lying resonances. In other words the
 $^{12}$C+$^{16}$O resonance energies are predicted from the
 quartet excitations of the 
$^{28}$Si. In particular, the projection was done by taking the intersection of
the quartet and cluster spectra in the superdeformed valley (in the second minimum of 
the energy-versus-deformation function, where the SD state corresponds to the ``ground''-band).
In order to characterize the breaking (or the goodness) of the MUSY
quantitatively, we have calculated the
\begin{equation}
sb=\frac{\sum_{i}|E^{exp}_i-E^{th}_i|}{\sum_{i}E^{exp}_i}
%\nonumber
\end{equation}
ratio. It turned out to be 13 \% for the spectrum of Figure 1
(including both the low- and the high-energy parts).      
When the resonances are also taken into account in the fitting procedure
(e.g. with a weight of 0.1 compared to the weight of 1.0 of the states in the 
well-established bands), a slightly better agreement of $sb = 12 \%$ can be obtained.
The low-energy bands 
(0(12,0)0$^+$, 1(14,1)1$^-$, 2(16,2)0$^+$, 4(20,4)0$^+$)
have single multiplicity in the shell-model space, therefore, the overlap of the
wavefunctions of their states in the quartet and cluster descriptions is 100\%.
(In the shell-model expansion of the cluster states there is only a single term.)

The  $^{24}$Mg+$^{4}$He cluster spectrum contains all the states
shown in Figure 1.
In the low-energy spectrum (lower part of Figure 1)
all the bands, except the
0(2,8)2$^+$, and 0(8,2)2$^+$
have single multiplicity in the shell-model.
As a consequence their wavefunctions are identical with those of the
$^{24}$Mg+$^{4}$He cluster configuration, as well as with those of the
$^{12}$C+$^{16}$O
clusterization, when it is allowed.

\begin{figure}[placement !]
\includegraphics[height=9.8cm,angle=0.]{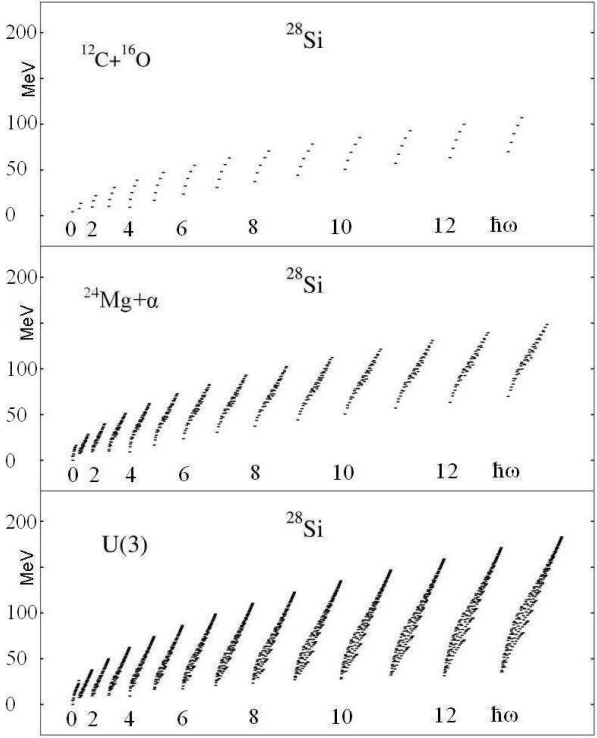}
\caption{ 
The landscape of the quartet and cluster band-heads in the 
 $^{28}$Si nucleus.
\label{fig:landscape}}
\end{figure}

For further illustration we show in Figure 2
the landscape of the bandhead-states in the 0-13 $\hbar \omega$
major shells for the quartet and cluster spectra.

As for the other possible binary clusterizations
(e.g.  $^{20}$Ne+$^{8}$Be)
of the $^{28}$Si nucleus the following can be said.
From the theoretical point of view they are available for
this kind of analysis, too, though technically some parts
might be more involved, due to the non-closed structure
(non SU(3) scalar nature) of the clusters.
At the same time, they are much less known from the
experimental side.

Further extension to non-alpha-like nuclei is also possible.
From the quartet side extra nucleons can be 
included when the semimicroscopic model is applied, like here
(as opposed to the phenomenologic quartet model),
since this approach is based on the nucleon degrees of freedom
\cite{quart}. 
The semimicroscopic algebraic cluster model allows the treatment
of the cluster with odd mass number, as well
\cite{odd},
due to the same reason.

In conclusion we can say that the  
semimicroscopic algebraic models are able to describe the quartet and cluster spectra 
in light nuclei in a unified framework. In particular: the multichannel dynamical symmetry gives
the cluster spectra from that of the quartet model by simple projections, therefore,
it has a very strong predictive power. In case of $^{28}$Si e.g. the high-lying spectrum of the
$^{12}$C+$^{16}$O clusterization is predicted from the low-lying quartet spectrum in remarkable 
agreement with the experimental observation.

%\section*{Acknowledgment}
\noindent
{\it This work was supported in part by the Hungarian Scientific 
Research Fund -  OTKA (Grant No K112962), and by the
International Collaboration of the
Al-Farabi Kazakh National University (Grant No: EP13). \\
%3106/GF4, and  1550/GF3).
}

\end{document}